\documentclass[11pt]{article}

\usepackage{times,latexsym,graphicx,amsmath,amsfonts,amsthm}
\usepackage{natbib}

\newtheoremstyle{localthm}
	{5pt} 
	{5pt} 
	{\sl} 
	{} 
	{\bf} 
	{{\rm.}} 
	{.7em} 
	{} 

\theoremstyle{localthm}

\newtheoremstyle{localrem}
	{5pt} 
	{5pt} 
	{\rm} 
	{} 
	{\bf} 
	{{\rm.}} 
	{.7em} 
	{} 

\theoremstyle{localrem}

\def\bs{\boldsymbol}

\def\ab{\bs{a}}
\def\bb{\bs{b}}
\def\fb{\bs{f}}
\def\zb{\bs{z}}

\def\lamb{\bs{\lambda}}

\def\fbhat{\hat{\bs{f}}}

\def\R{\mathbb{R}}

\def\FF{\mathcal{F}}
\def\PP{\mathcal{P}}

\def\argmin{\mathop{\rm arg\,min}}

\begin{document}

\addtolength{\baselineskip}{+.2\baselineskip}

\title{A New Algorithm for Totally Positive Approximations}

\author{Philip A.\ Stange and Lutz D{\"u}mbgen\\
University of Bern}

\date{\today}

\maketitle

\paragraph{Abstract.}
We revisit the problem of approximating a bivariate distribution with finite support by another such distribution which is totally positive of order two (TP2). Approximation is meant in a maximum likelihood sense.



\section{Introduction}
\label{sec:Introduction}

Distributional regression for random variables $(X,Y) \in \R\times \R$ means estimating the conditional distributions $\mathcal{L}(Y \,|\, X = x)$, $x \in \R$. One possibility to estimate these conditional distributions nonparametrically is to impose a monotonicity constraint, see \cite{Duembgen_2024} for a general introduction to this approach. Specifically, one can estimate the conditional distributions under the constraint that $\mathcal{L}(Y \,|\, X = x_1) \le_{\rm lr} \mathcal{L}(Y \,|\, X = x_2)$ whenever $x_1 < x_2$, where $\le_{\rm lr}$ denotes the likelihood ratio order, see \cite{Duembgen_Moesching_2023}. As shown in the latter paper, this constraint is equivalent to assuming that the joint distribution $\mathcal{L}(X,Y)$ is totally positive of order two (TP2). This means that for arbitrary Borel sets $A_1 < A_2$ and $B_1 < B_2$ (elementwise) in $\R$,
\[
	\mathrm{P}(X \in A_1, Y \in B_1) \mathrm{P}(X \in A_2, Y \in B_2) \
	\ge \ \mathrm{P}(X \in A_1, Y \in B_2) \mathrm{P}(X \in A_2, Y \in B_1) .
\]
Hence, the new question is to estimate a distribution $R$ on $\R\times \R$ under the constraint that it is TP2.

While \cite{Duembgen_Moesching_2023} proposed a minimum distance estimator based on a bivariate Kuiper distance, \cite{Moesching_Duembgen_2024} used an empirical likelihood approach in the spirit of \cite{Owen_2001}: Starting from a data set with observations $(X_\ell,Y_\ell) \in \R \times \R$, $\ell \in [N] := \{1,2,\ldots,N\}$, we want to approximate their empirical distribution by a distribution $\hat{R}$ on the grid
\[
	\{X_1,\ldots,X_n\} \times \{Y_1,\ldots,Y_n\}
\]
under the constraint that $\hat{R}$ is totally positive of order two (TP2). This means that the point mass function $\hat{r}$ of $\hat{R}$ satisfies
\[
	\hat{r}(x_1,y_1) \hat{r}(x_2,y_2) \ \ge \ \hat{r}(x_1,y_2) \hat{r}(x_2,y_1)
\]
for arbitrary $x_1 < x_2$ in $\{X_1,\ldots,X_n\}$ and $y_1 < y_2$ in $\{Y_1,\ldots,Y_n\}$. Approximation is meant in the sense of maximum empirical likelihood \citep{Owen_2001}, that is, $\hat{r}$ maximizes the log-likelihood
\[
	\sum_{\ell=1}^N \log \hat{r}(X_\ell,Y_\ell) .
\]

More generally, for given points $x_1 < \cdots < x_m$ and $y_1 < \cdots < y_n$, let $R$ be a finite measure on $\{x_1,\ldots,x_m\} \times \{y_1,\ldots,y_n\}$ with point masses denoted by
\[
	w_{ij} \ := \ R(\{x_i\} \times \{y_j\}) .
\]
We assume that all row sums $w_{i+}$, $i \in [m]$, and all column sums $w_{+j}$, $j \in [n]$, of $\bs{w} = (w_{ij})_{i,j}$ are strictly positive. As shown by \cite{Moesching_Duembgen_2024}, there exists a unique minimizer $\fbhat$ of the loss function
\[
	L(\fb) \ := \ \sum_{i=1}^m \sum_{j=1}^n (- w_{ij} \log f_{ij} + f_{ij} - w_{ij})
\]
over the set $\FF$ of all $\fb \in [0,\infty)^{m\times n}$ such that
\begin{equation}
\label{eq:TP2}
	f_{i_1j_1} f_{i_2,j_2} \ \ge \ f_{i_1j_2} f_{i_2j_1}
	\quad \text{for} \ 1 \le i_1 < i_2 \le m , 1 \le j_1 < j_2 \le n .
\end{equation}
The minimizer $\fbhat$ satisfies automatically the equations $\hat{f}_{i+} = w_{i+}$ for $i \in [m]$ and $\hat{f}_{+j} = w_{+j}$ for $j \in [n]$.

In the specific setting at the beginning, $x_1 < \ldots < x_m$ are the different elements of the set $\{X_1,X_2,\ldots,X_N\}$, $y_1 < \ldots < y_n$ are the different elements of $\{Y_1,Y_2,\ldots,Y_N\}$, and $w_{ij}$ is the number of $\ell \in [N]$ such that $X_\ell = x_i$ and $Y_\ell = y_j$. Then the maximum empirical likelihood estimator $\hat{r}$ is given by $\hat{r}(x_i,y_j) = \hat{f}_{ij}/N$, and the resulting log-likelihood is equal to $- L(\fbhat)$.

\cite{Moesching_Duembgen_2024} presented an iterative algorithm for the computation of $\fbhat$ for arbitrary weight matrices $\bs{w} \in [0,\infty)^{m\times n}$ with strictly positive row and column sums. But their algorithm is rather slow for large values of $m$ and $n$. Our goal is to simplify and accelerate that algorithm. In particular, for the iterative procedure we come up with, a single iteration is based on $n-1$ shape-constrained Newton steps on $m$- or lower-dimensional submanifolds of the parameter space, and it requires $O(mn)$ steps and memory.

\section{The algorithm}
\label{sec:algorithm}

\paragraph{Calibration.}
If $\fb \in \FF$ is a candidate for $\fbhat$, it can be improved by row-wise and column-wise calibration. Precisely,
\[
	L \bigl( (f_{ij} w_{i+}^{}/f_{i+})_{i,j} \bigr) \ \le \ L(\fb)
\]
with equality if and only if $f_{i+} = w_{i+}^{}$ for all $i \in [m]$. Indeed, for any $\ab \in \R^m$, the matrix $\fb(\ab) = \bigl( f_{ij} e^{a_i} \bigr)_{i,j}$ satisfies
\[
	L(\fb(\ab)) \
	= \ L(\fb) + \sum_{i=1}^m \bigl( - w_{i+}^{} a_i + (e^{a_i} - 1) f_{i+} \bigr) .
\]
This is minimal in $\ab$ if and only if $a_i = \log(f_{i+}/w_{i+}^{})$ for $i \in [m]$, and the resulting matrix $\fb(\ab)$ can be written as $\bigl( f_{ij} w_{i+}^{}/f_{i+} \bigr)_{i,j}$.

Analogously,
\[
	L \bigl( (f_{ij} w_{+j}/f_{+j})_{i,j} \bigr) \ \le \ L(\fb)
\]
with equality if and only if $f_{+j} = w_{+j}$ for all $j \in [n]$.

\paragraph{The special case of $n = 2$ or $m = 2$.}
It is instructive to look at the special case $n = 2$ first. If $\fb \in \FF$ is row-wise calibrated, it may be written as
\[
	f_{ij} \ = \ w_{i+}^{} (1_{[j = 1]} (1 - \lambda_i) + 1_{[j = 2]} \lambda_i)
\]
with a vector $\lamb$ in the set $[0,1]^m_{\uparrow}$ of vectors with $m$ non-decreasing components in $[0,1]$. Then,
\begin{align*}
	L(\fb) \
	&= \ \sum_{i=1}^m \bigl( - w_{i+}^{} \log w_{i+}^{} - w_{i1}^{} \log(1 - \lambda_i)
		- w_{i2}^{} \log \lambda_i + w_{i+}^{} \bigr) \\
	&= \ \sum_{i=1}^m w_{i+}^{} \log(e/w_{i+}^{})
		- \sum_{i=1}^m w_{i+}^{} \bigl( (1 - p_i) \log(1 - \lambda_i) + p_i \log \lambda_i \bigr)
\end{align*}
with $p_i := w_{i2}^{}/w_{i+}^{}$. This is just a constant plus a weighted negative log-likelihood function for isotonic binary regression, see \cite{Robertson_etal_1988}. In particular, the optimal choice for $\lamb$ is the unique minimizer
\[
	\widehat{\lamb} \
	:= \ \argmin_{\lamb \in [0,1]^m_{\uparrow}} \,
		\sum_{i=1}^m w_{i+}^{} (\lambda_i - p_i)^2 .
\]
It can be computed via the pool-adjacent-violators algorithm (PAVA) in $O(m)$ steps.

Analogously one can compute $\fbhat$ in the special case of $m = 2$ by means of the PAVA.

\paragraph{The support of $\fbhat$.}
From now on let us focus on the situation in which $m$ and $n$ are at least $3$. With $[m\times n] := [m] \times [n]$, let $\PP$ consist of all index pairs $(i,j) \in [m\times n]$ such that there exist indices $i_1 \le i \le i_2$ in $[m]$ and $j_1 \le j \le j_2$ in $[n]$ with $w_{i_1j_2} > 0$ and $w_{i_2j_1} > 0$. When miminimizing $L(\fb)$ over all $\fb \in \FF$, it suffices to consider matrices $\fb$ such that
\begin{equation}
\label{eq:suPP}
	\{(i,j) \in [m\times n] : f_{ij} > 0\} \ = \ \PP ,
\end{equation}
see Lemma~3.1 of \cite{Moesching_Duembgen_2024}. In what follows we write $\FF_\PP$ for the set of all $\fb \in \FF$ satisfying \eqref{eq:suPP}.

\paragraph{A particular update step.}
Suppose that $\fb \in \FF_\PP$ is a candidate for $\fbhat$ which is row-wise calibrated. For a fixed index $k \in [m-1]$, let
\[
	I(k) \ := \ \bigl\{ i \in [m] : (i,k), (i,k+1) \in \PP \bigr\} .
\]
Suppose that $I(k) \ne \emptyset$. Now we consider arbitrary vecors $\ab, \bb \in \R^{I(k)}$ and define $\fb(\ab,\bb)$ via
\[
	f_{ij}(\ab,\bb) \ := \ f_{ij} \cdot
		\begin{cases}
			e^{a_i} & \text{if} \ i \in I(k) \ \text{and} \ j \le k , \\
			e^{a_i + b_i} & \text{if} \ i \in I(k) \ \text{and} \ j > k , \\
			1 & \text{else} .
		\end{cases}
\]
To compare $L(\fb(\ab,\bb))$ with $L(\fb)$, we set
\[
	\bar{p}_i \ := \ w_{i+}^{-1} \sum_{j=1}^{k} w_{ij} , \quad
	p_i \ := \ w_{i+}^{-1} \sum_{j=k+1}^n w_{ij} , \quad
	\bar{g}_i \ := \ w_{i+}^{-1} \sum_{j=1}^{k} f_{ij} , \quad
	g_i \ := \ w_{i+}^{-1} \sum_{j=k+1}^n f_{ij}
\]
for $i \in [m]$. Obviously, $\bar{p}_i + p_i = 1$, and $\bar{g}_i + g_i = 1$ as well, because $f_{i+} = w_{i+}^{}$ by assumption. Moreover, $\bar{g}_i$ and $g_i$ are strictly positive for $i \in I(k)$. Now we can write
\begin{align*}
	L(\fb(\ab,\bb)) \
	&= \ L(\fb) + \sum_{i \in I(k)} \sum_{j=1}^n
		\bigl( - w_{ij} (a_i + 1_{[j > k]} b_i)
			+ f_{ij} \bigl( e^{a_i} (1 + 1_{[j > k]} e^{b_i}) - 1 \bigr) \bigr) \\
	&= \ L(\fb) + \sum_{i \in I(k)} w_{i+}^{}
		\bigl( - a_i - p_i b_i
			+ e^{a_i} (\bar{g}_i + g_i e^{b_i}) - 1 \bigr) .
\end{align*}
For fixed $\bb$, this is minimal in $\ab$ if and only if $\ab = \ab(\bb)$ with
\[
	a_i(\bb) \
	= \ - \log(\bar{g}_i + g_i e^{b_i}) .
\]
This leads to
\[
	\Lambda(\bb) \ := \ L \bigl( \fb(\ab(\bb),\bb) \bigr) - L(\fb) \
	= \ \sum_{i\in I(k)} w_{i+}^{}
		\bigl( \log(\bar{g}_i + g_i e^{b_i}) - p_i b_i \bigr) .
\]

If $I(k)$ is a singleton $\{i(k)\}$, then $\fb(\ab,\bb) \in \FF_\PP$ for any choice of $\ab,\bb$. The target functional $\Lambda(\bb)$ is minimal if and only if
\[
	b_{i(k)} \ = \ \log \Bigl( \frac{p_{i(k)}\bar{g}_{i(k)}}{\bar{p}_{i(k)} g_{i(k)}} \Bigr) ,
\]
and the corresponding $\ab = \ab(\bb)$ is given by
\[
	a_{i(k)} \ = \ \log \Bigl( \frac{\bar{p}_{i(k)}}{\bar{g}_{i(k)}} \Bigr) .
\]
Hence, the optimal matrix $\fb(\ab,\bb) \in \FF_\PP$ is given by
\[
	f_{ij}(\ab,\bb) \ = \ f_{ij} \cdot
		\begin{cases}
			\bar{p}_{i(k)}/\bar{g}_{i(k)} & \text{if} \ i = i(k) \ \text{and} \ j \le k , \\
			p_{i(k)}/g_{i(k)} & \text{if} \ i = i(k) \ \text{and} \ j > k , \\
			1 & \text{else} .
		\end{cases}
\]

If $I(k)$ has at least two elements, things get a bit more complicated. Note that
\begin{align*}
	\log(\bar{g}_i + g_i e^{b_i}) \
		= \ \log(1 + g_i (e^{b_i} - 1)) \
	&= \ g_i (e^{b_i} - 1) - g_i^2 (e^{b_i} - 1)^2/2 + O(b_i^3) \\
	&= \ g_i b_i + (g_i - g_i^2) b_i^2/2 + O(b_i^3) \\
	&= \ g_i b_i + g_i \bar{g}_i b_i^2/2 + O(b_i^3)
\end{align*}
as $b_i \to 0$. Hence, as $\bb \to \bs{0}$,
\[
	\Lambda(\bb) \
	= \ \frac{1}{2} \sum_{i\in I(k)} w_{i+}^{} g_i \bar{g}_i
		\Bigl( b_i^2 - 2 b_i \frac{p_i - g_i}{g_i \bar{g}_i} \Bigr)
		+ O(\|\bb\|^3) .
\]
Consequently, to get a promising candidate for $\bb$, we minimize
\[
	\sum_{i \in I(k)} w_{i+}^{} g_i \bar{g}_i
		\Bigl( b_i - \frac{p_i - g_i}{g_i \bar{g}_i} \Bigr)^2
\]
over all $\bb \in \R^{I(k)}$ such that $\fb(\ab(\bb),\bb) \in \FF_\PP$. This is equivalent to the requirement that
\[
	\bigl( b_i + \log(f_{i,k+1}/f_{ik}) \bigr)_{i \in I(k)} \
	\in \ \R^{I(k)}_{\uparrow} ,
\]
the cone of vectors in $\R^{I(k)}$ with non-decreasing components. Consequently, we choose
\[
	\bb = \bb(\fb) \
	:= \ \zb(\fb) - \bigl( \log(f_{i,k+1}/f_{ik}) \bigr)_{i \in I(k)}
\]
with the unique minimizer
\[
	\zb(\fb) \ := \ \argmin_{\zb \in \R^{I(k)}_\uparrow} \,
		\sum_{i \in I(k)} w_{i+}^{} g_i \bar{g}_i \Bigl(
		z_i - \frac{p_i - g_i}{g_i \bar{g}_i} - \log (f_{i,k+1}/f_{ik})
		\Bigr)^2 .
\]

There is no guarantee that the new candidate $\fb(\ab(\bb),\bb)$ is better than $\fb$. Thus we introduce a step size correction. For $t \in [0,1]$ we consider the improvement function
\[
	h(t) \
	:= \ - \Lambda(t\bb) \
	= \ \sum_{i \in I(k)} w_{i+}^{}
		\bigl( t p_i b_i - \log(\bar{g}_i + g_i e^{tb_i}) \bigr)
\]
and note that $h$ is concave with
\[
	h'(0) \ = \ \sum_{i\in I(k)} w_{i+}^{} (p_i - g_i) b_i .
\]
Thus we leave $\fb$ unchanged if $h'(0)$ is smaller than a given small threshold $\delta_o > 0$, e.g.\ $\delta_o = 10^{-7}$. Otherwise we replace $\fb$ with $\fb(\ab(2^{-\kappa}\bb),2^{-\kappa}\bb)$, where $\kappa$ is the smallest nonnegative integer such that $h(2^{-\kappa}) \ge 2^{-\kappa} h'(0)/3$. In practice, we consider only $\kappa \in \{0,\ldots,\kappa_o\}$ for a given integer $\kappa_o$, e.g.\ $\kappa_o = 20$. If we reach this integer $\kappa_o$, then $\fb$ is replaced with $\fb(\ab(2^{-\kappa_o}\bb),2^{-\kappa_o}\bb)$ if $h(2^{-\kappa_o}) > 0$.

\paragraph{Combining $n-1$ updates as one iteration.}
Our goal is to perform the previous updates for $k = 1,2,\ldots,n-1$ such that these $n-1$ updates require $O(mn)$ steps and memory in total. Starting from a row-wise calibrated candidate $\fb = \fb^{(1)}$, let $\fb^{(k+1)}$ be its updated version after performing the $k$-th update, so $\fb^{(n)}$ is the final new candidate. The main ingredients for the $k$-th update are the vectors
\[
	\bs{p}^{(k)} \ = \ \Bigl( w_{i+}^{-1} \sum_{j > k} w_{ij}^{} \Bigr)_{i=1}^m ,
	\quad
	\bar{\bs{p}}^{(k)} \ = \ \Bigl( w_{i+}^{-1} \sum_{j \le k} w_{ij}^{} \Bigr)_{i=1}^m ,
\]
\[
	\bs{g}^{(k)} \ = \ \Bigl( w_{i+}^{-1} \sum_{j > k} f_{ij}^{(k)} \Bigr)_{i=1}^m ,
	\quad
	\bar{\bs{g}}^{(k)} \ = \ \Bigl( w_{i+}^{-1} \sum_{j \le k} f_{ij}^{(k)} \Bigr)_{i=1}^m
\]
and $\bs{\delta}^{(k)} \in \R^m$ with components
\[
	\delta_i^{(k)} \ = \ \begin{cases}
		\log(f_{i,k+1}/f_{ik}) & \text{if} \ i \in I(k) , \\
		0 & \text{else} .
	\end{cases}
\]
We use deliberately $f_{i,k+1}/f_{ik}$ instead of $f_{i,k+1}^{(k)}/f_{ik}^{(k)}$ as explained later. Computing the four vectors $\bs{p}^{(1)}, \bar{\bs{p}}^{(1)}, \bs{g}^{(1)}, \bar{\bs{g}}^{(1)}$ and the $n-1$ vectors $\bs{\delta}^{(k)}$, $k \in [n-1]$, requires $O(mn)$ steps.

Updating $\fb^{(k)}$ amounts to replacing it with $\fb^{(k+1)}$ with components
\[
	f_{ij}^{(k+1)} \ = \ f_{ij}^{(k)} \exp \bigl( a_i^{(k)} + 1_{[j > k]} b_i^{(k)} \bigr)
\]
for suitable vectors $\ab^{(k)}, \bb^{(k)} \in \R^m$ such that $a_i^{(k)} = b_i^{(k)} = 0$ for $i \not\in I(k)$, and the determination of the latter two vectors requires $O(m)$ steps.

If we would really update the full matrix $\fb^{(k)}$, the total running time would become $O(mn^2)$, simply because there are $mn$ components to be updated for each $k$. But this can be avoided by means of suitable bookkeeping. First of all, note that replacing $\fb$ with $\fb^{(k')}$ for any $k' \le k$ would have no impact on the components of $\bs{\delta}^{(\ell)}$ for any $\ell > k$. This is why we defined $\bs{\delta}^{(k)}$ with the original candidate $\fb$. The vectors $\bs{p}^{(k+1)}$ and $\bar{\bs{p}}^{(k+1)}$ can be obtained via
\begin{align*}
	p_i^{(k+1)} \
	&= \ p_i^{(k)} - w_{i+}^{-1} w_{i,k+1}^{} , \\
	\bar{p}_i^{(k+1)} \
	&= \ \bar{p}_i^{(k)} - w_{i+1}^{-1} w_{i,k+1}^{}
\end{align*}
in $O(m)$ steps. The next vectors $\bs{g}^{(k+1)}$ and $\bar{\bs{g}}^{(k+1)}$ can be represented as follows:
\begin{align*}
	g_i^{(k+1)} \
	&= \ (g_i^{(k)} - w_{i+}^{-1} f_{i,k+1}^{(k)}) \exp(a_i^{(k)} + b_i^{(k)})  \\
	&= \ g_i^{(k)} \exp(a_i^{(k)} + b_i^{(k)})
		- w_{i+}^{-1} f_{i,k+1} \exp \Bigl( \sum_{k'\le k} (a_i^{(k')} + b_i^{(k')}) \Bigr) , \\
	\bar{g}_i^{(k+1)} \
	&= \ \bar{g}_i^{(k)} \exp(a_i^{(k)}) + w_{i+}^{-1} f_{i,k+1}^{(k)} \exp(a_i^{(k)} + b_i^{(k)}) \\
	&= \ \bar{g}_i^{(k)} \exp(a_i^{(k)})
		+ w_{i+}^{-1} f_{i,k+1} \exp \Bigl( \sum_{k'\le k} (a_i^{(k')} + b_i^{(k')}) \Bigr) .
\end{align*}
Consequently, if we define $\bs{A}^{(0)} := \bs{B}^{(0)} := \bs{0} \in \R^m$ and
\[
	\bs{A}^{(k)} \ := \ \Bigl( \sum_{k' \le k} a_i^{(k')} \Bigr)_{i=1}^m
	\quad\text{and}\quad
	\bs{B}^{(k)} \ := \ \Bigl( \sum_{k' \le k} b_i^{(k')} \Bigr)_{i=1}^m ,
\]
then $\bs{A}^{(k)}$ and $\bs{B}^{(k)}$ can be obtained via the recursions $A_i^{(k)} = A_i^{(k-1)} + a_i^{(k)}$, $B_i^{(k)} = B_i^{(k-1)} + b_i^{(k)}$, and in case of $k < n-1$, the new vectors $\bs{g}^{(k+1)}$ and $\bar{\bs{g}}^{(k+1)}$ are computed as
\begin{align*}
	g_i^{(k+1)} \
	&= \ g_i^{(k)} \exp(a_i^{(k)} + b_i^{(k)})
		- w_{i+}^{-1} f_{i,k+1} \exp(A_i^{(k)} + B_i^{(k)}) , \\
	\bar{g}_i^{(k+1)} \
	&= \ \bar{g}_i^{(k)} \exp(a_i^{(k)})
		+ w_{i+}^{-1} f_{i,k+1} \exp(A_i^{(k)} + B_i^{(k)}) .
\end{align*}
For a fixed $k$, all these recursions require $O(m)$ steps in total. And the final candidate $\fb^{(n)}$ can be obtained in $O(mn)$ steps via the representation
\[
	f_{ij}^{(n)} \ = \ f_{ij}^{} \exp \bigl( A_i^{(n-1)} + B_i^{(j-1)} \bigr) ,
\]
provided that the vectors $\bs{B}^{(j-1)}$, $2 \le j \le n$, have been stored properly, which requires $O(mn)$ additional memory.

\paragraph{The complete algorithm.}
The complete algorithm repeats the previous iteration until the decrease in the target function $L$ is no larger than a given small constant $\delta > 0$, e.g.\ $\delta = 10^{-3}$. Numerical experiments also revealed that ``rotating'' the triplets $(\bs{w},\fb,\bs{p})$ with $\bs{p} = (1_{\PP}(i,j))_{i,j}$ is beneficial. That is, before iteration $r$, we replace $(\bs{w},\fb,\bs{p})$ with the triplet
\[
	\bigl( T_{v(r)}(\bs{w}), T_{v(r)}(\fb), T_{v(r)}(\bs{p}) \bigr) ,
\]
apply the iteration described above to the latter and retransform the updated matrix $T_{v(r)}(\fb)_{\rm new}$ to get $\fb_{\rm new} = T_{v(r)}(T_{v(r)}(\fb)_{\rm new})$. Here, $v(r) = (r-1) \,\mathrm{mod}\, 4$, and
\[
	T_0(\bs{A}) \ := \ \bs{A} , \quad
	T_1(\bs{A}) \ := \ \bs{A}^\top , \quad
	T_2(\bs{A}) \ := \ (A_{m+1-i,n+1-j})_{i,j} , \quad
	T_3(\bs{A}) \ := \ T_2(\bs{A})^\top
\]
for $\bs{A} \in \R^{m\times n}$. Note that $T_v(\fb)$ is still TP2 and $T_v(T_v(\bs{A})) = \bs{A}$ for $v=0,1,2,3$.

\section{A numerical example}
\label{sec:example}

\begin{figure}
\includegraphics[width=0.49\textwidth]{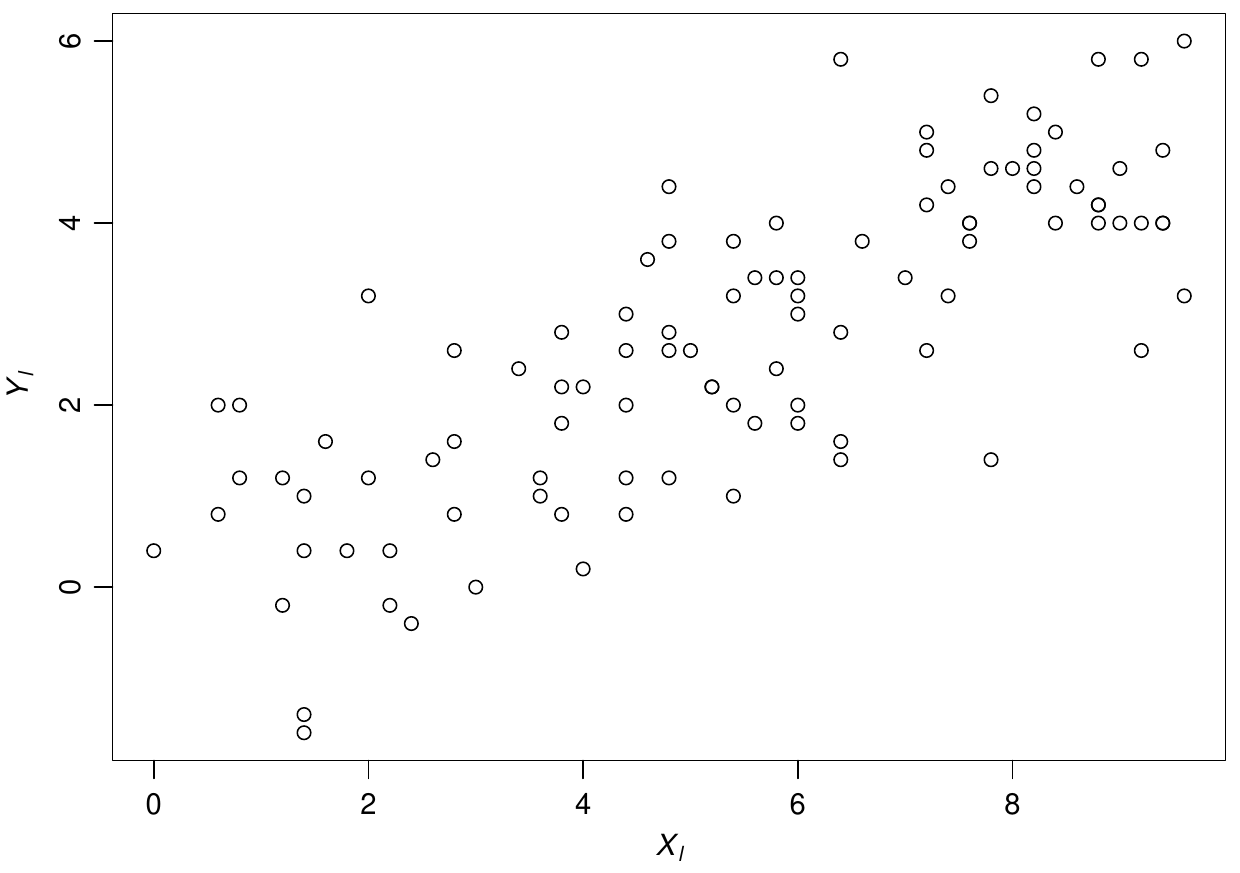}
\hfill
\includegraphics[width=0.49\textwidth,height=0.24\textheight]{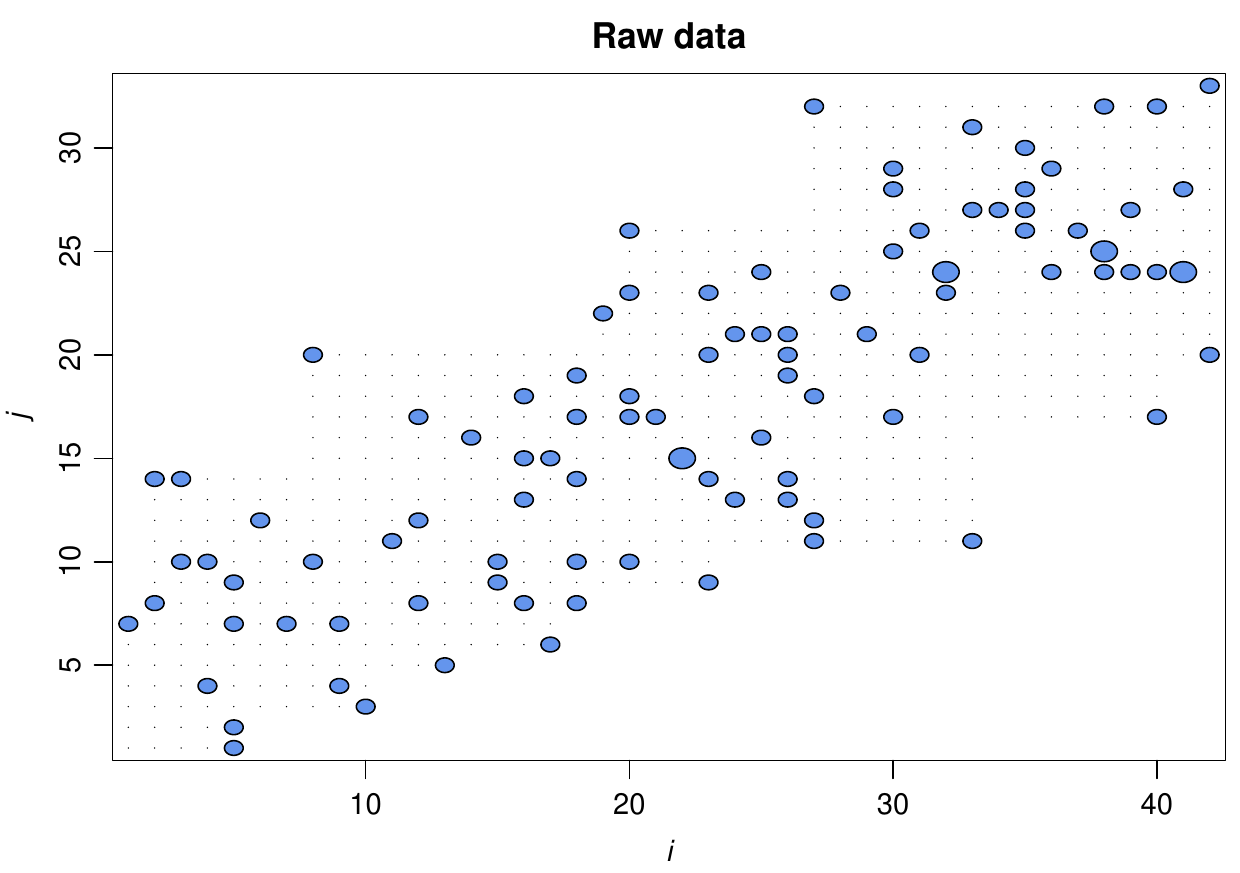}
\caption{Simulated data (left) and corresponding weight matrix (right).}
\label{fig:RawData}
\end{figure}

To illustrate the algorithm, we simulated a data set with $N = 100$ independent identically distributed observations $(X_\ell,Y_\ell)$, $\ell \in [100]$; see the left panel of Figure~\ref{fig:RawData}. The right panel depicts the corresponding weight matrix $\bs{w} \in \R^{42 \times 33}$. The area of a circle at $(i,j)$ is proportional to $w_{ij}$.

\begin{figure}
\includegraphics[width=0.49\textwidth,height=0.24\textheight]{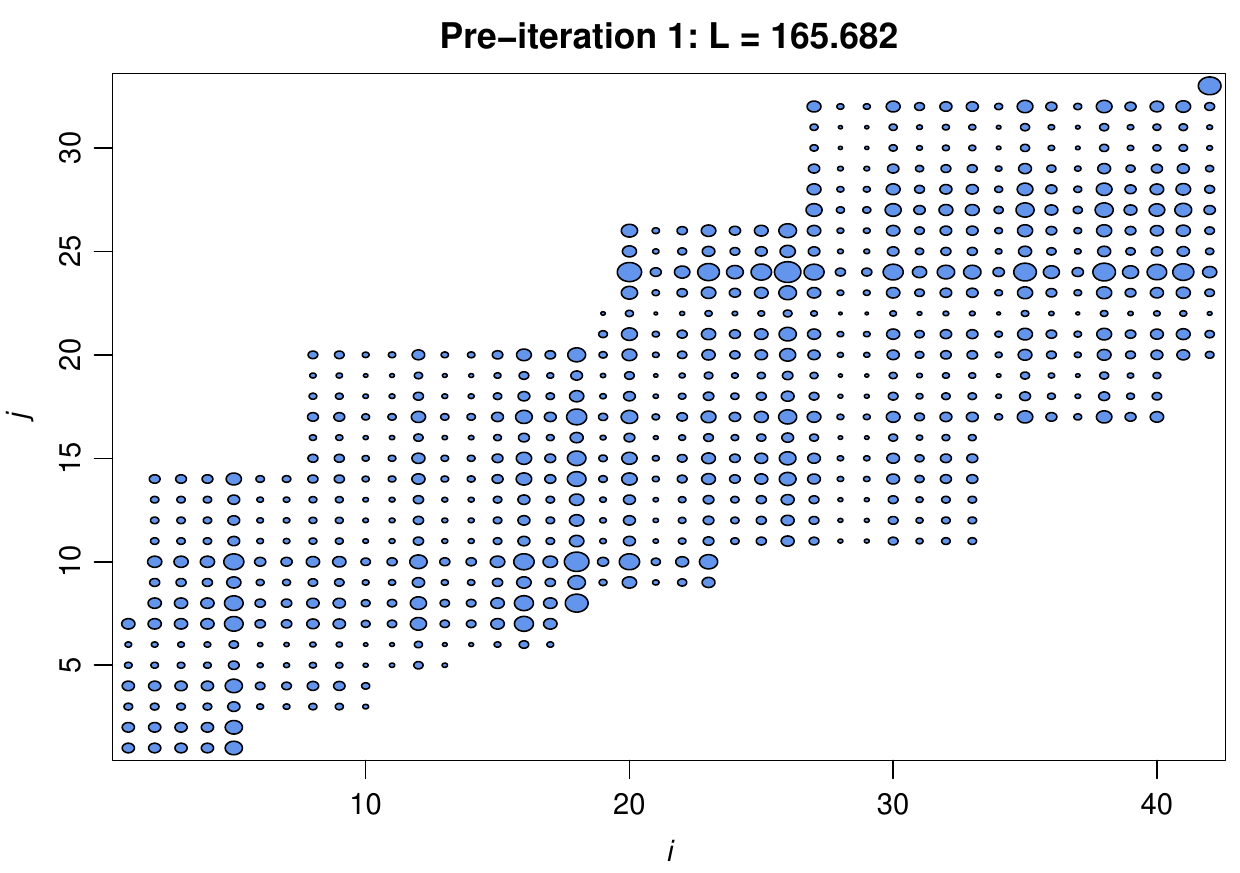}
\hfill
\includegraphics[width=0.49\textwidth,height=0.24\textheight]{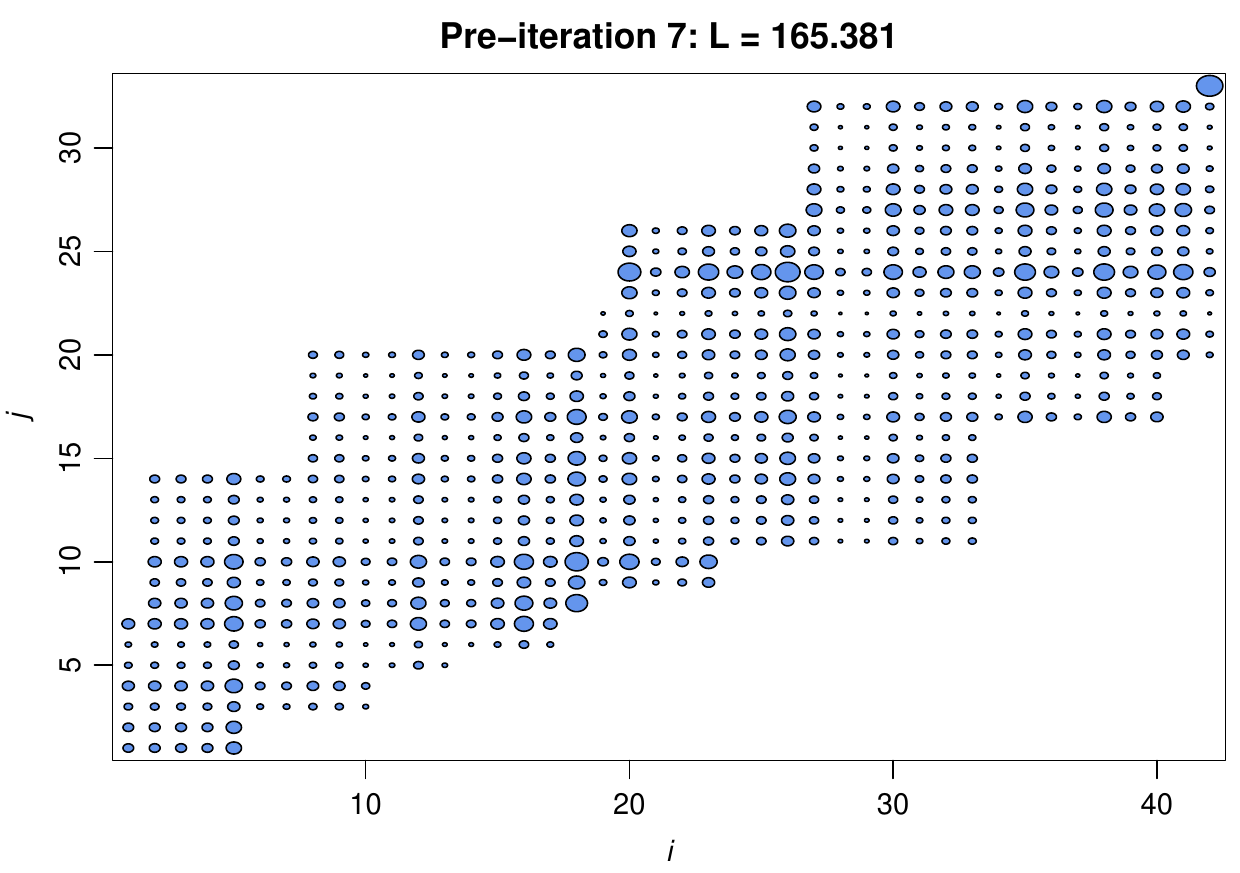}
\caption{Candidate $\bs{f}$ after one and seven pre-iterations.}
\label{fig:PreIter}
\end{figure}

Starting with $\bs{f} = \bigl( 1_{[(i,j) \in \PP]}^{} \bigr)_{i,j}$, as a first pre-iteration, we applied a row-wise calibration and a column-wise calibration of $\bs{f}$. This pre-iteration was repeated until the decrease of $L(\bs{f})$ was less than $10^{-4}$, which happened after seven pre-iterations. Figure~\ref{fig:PreIter} shows that candidate $\bs{f}$ after the first and after the last pre-iteration. Note that $\bs{f}$ still satisfies the condition that $f_{ij} f_{i+1,j+1} = f_{i,j+1} f_{i+1,j}$ whenever $(i,j+1), (i+1,j) \in \PP$.

\begin{figure}
\includegraphics[width=0.49\textwidth,height=0.24\textheight]{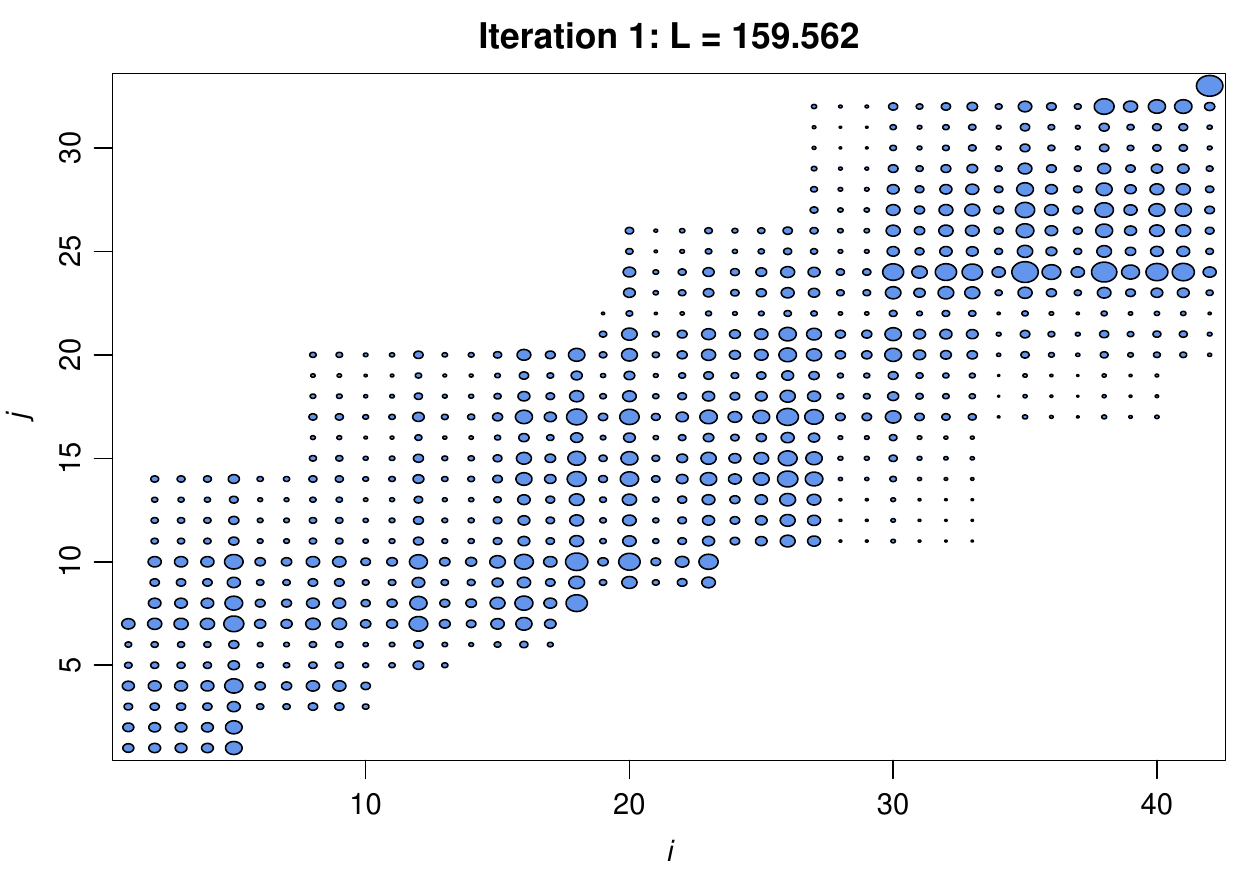}
\hfill
\includegraphics[width=0.49\textwidth,height=0.24\textheight]{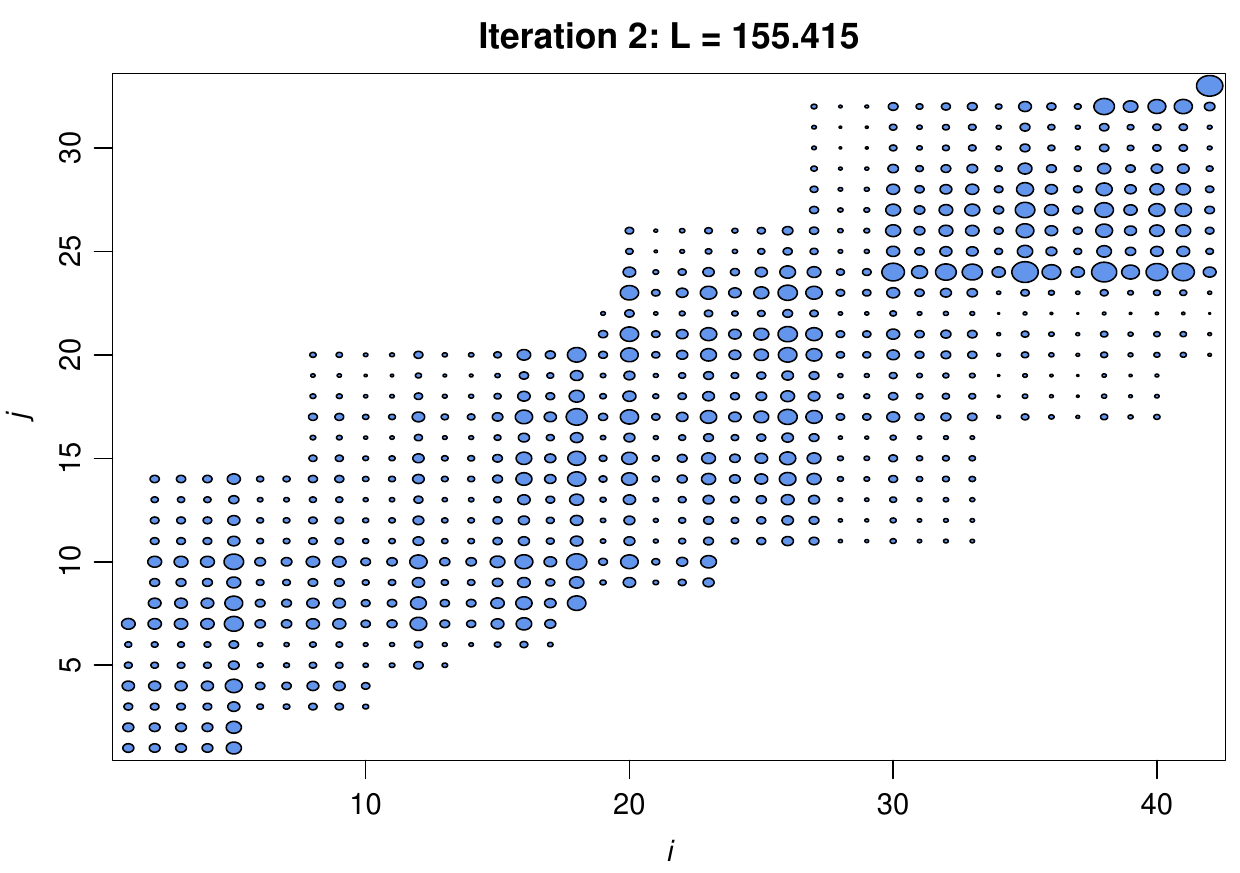}

\includegraphics[width=0.49\textwidth,height=0.24\textheight]{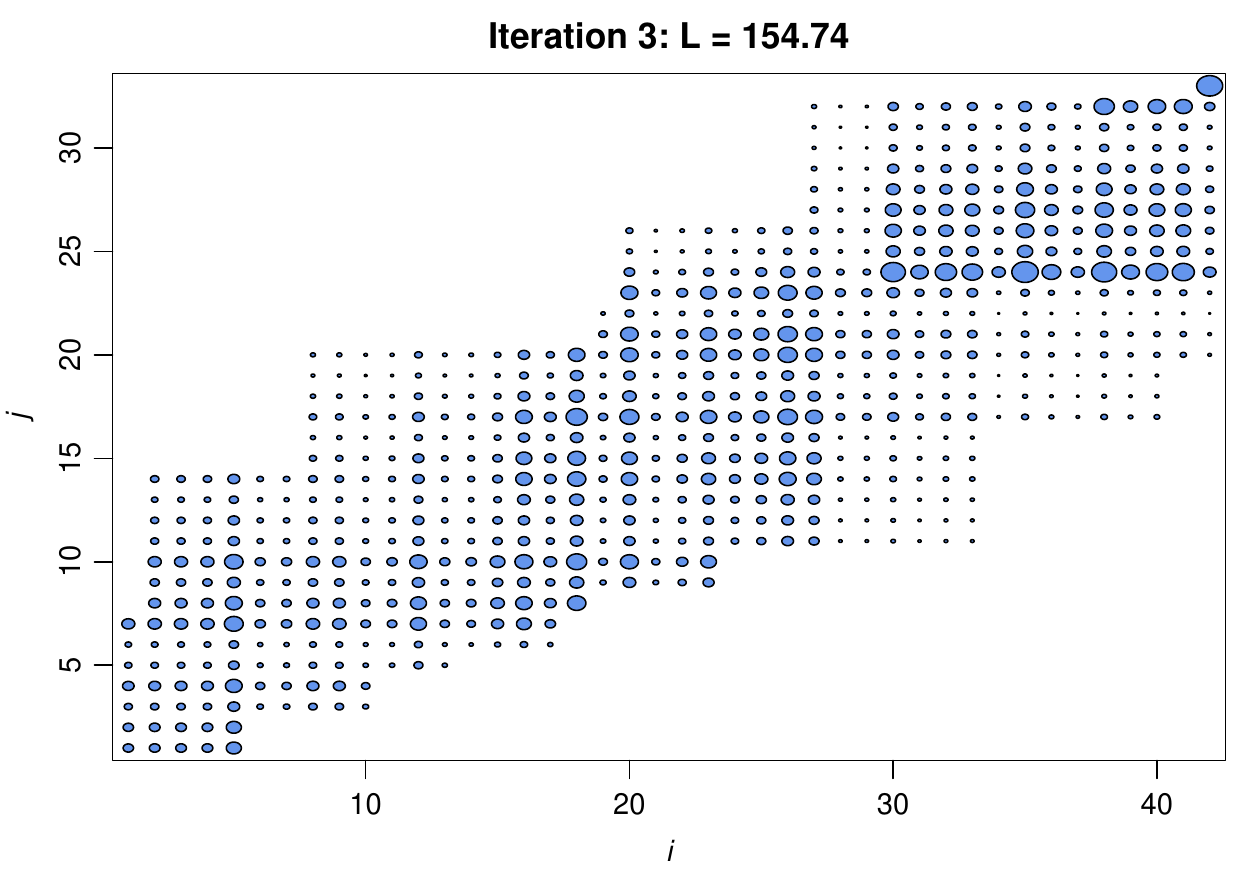}
\hfill
\includegraphics[width=0.49\textwidth,height=0.24\textheight]{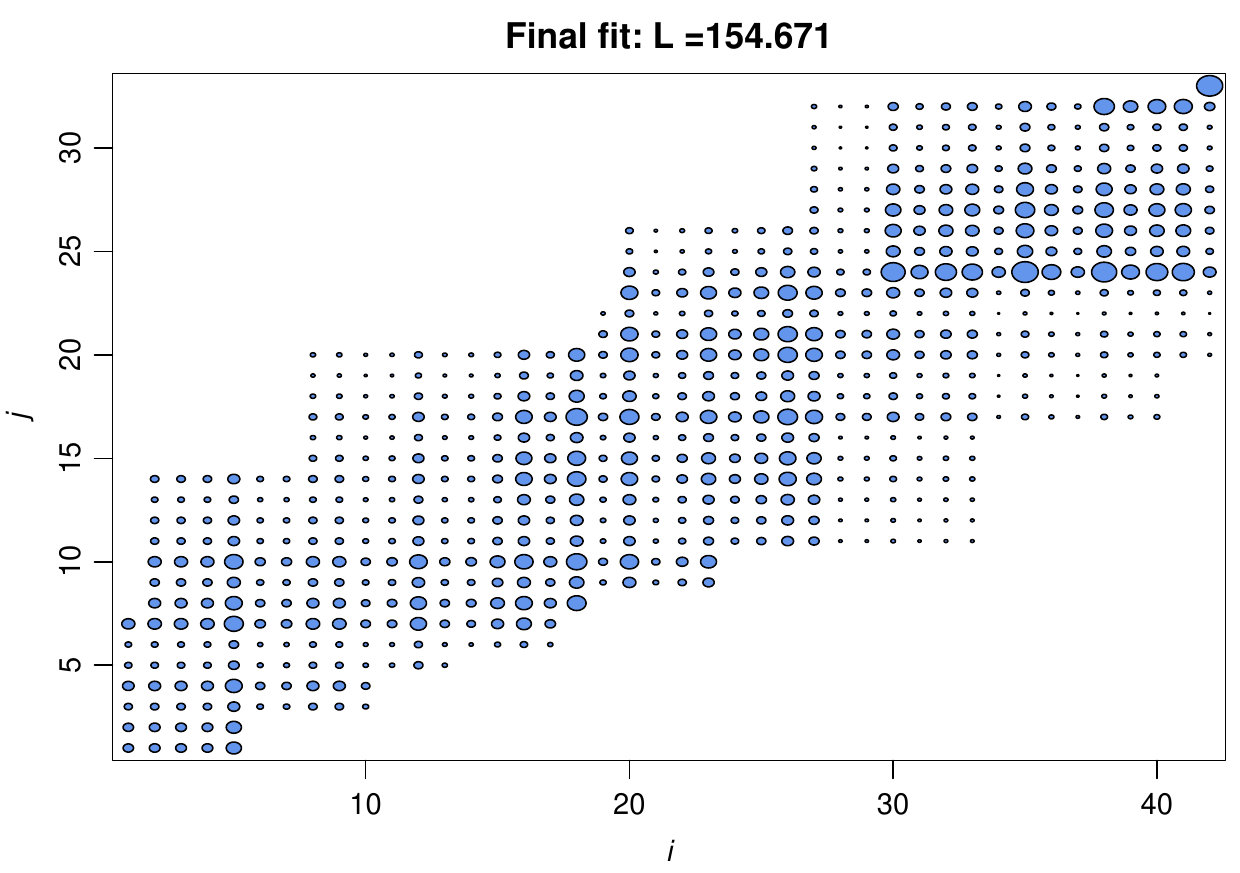}
\caption{Candidate $\bs{f}$ after one, two, three and eight iterations.}
\label{fig:Iter}
\end{figure}

After these pre-iterations, we started the full iterations described in the previous section and, again, stopped this process when the decrease of $L(\bs{f})$ was less than $10^{-4}$. This happened after eight iterations. Figure~\ref{fig:Iter} shows the resulting candidate $\bs{f}$ after one, two and three iterations and the final result.

\paragraph{Acknowledgement.}
This work was supported by the Swiss National Science Foundation.


\end{document}